\documentclass[a4paper,11pt]{article}
\usepackage{jcappub}
\usepackage[T1]{fontenc}
\usepackage[english]{babel}
\usepackage[utf8]{inputenc}
\usepackage[colorinlistoftodos, color=green!40, prependcaption]{todonotes}
\usepackage{orcidlink}
\usepackage{amsthm}
\usepackage{mathtools}
\usepackage{physics}
\usepackage{xcolor}
\usepackage{graphicx}
\usepackage{adjustbox}
\usepackage{placeins}
\usepackage[T1]{fontenc}
\usepackage{lipsum}
\usepackage{csquotes}
\usepackage{hyperref}
\usepackage{subfigure}
\usepackage{lipsum}
\usepackage[normalem]{ulem}

\newcommand{\GC}{\textcolor{black}}

\title{Low-Energy Supernovae Bounds on Sterile Neutrinos }
\author[a]{Garv Chauhan\,\orcidlink{0000-0002-8129-8034}}
    \emailAdd{gchauhan@vt.edu}
\author[a,b]{Shunsaku Horiuchi\,\orcidlink{0000-0001-6142-6556}}
    \emailAdd{horiuchi@vt.edu}
    \affiliation[a]{Center for Neutrino Physics, Department of Physics, Virginia Tech, Blacksburg, VA 24061, USA}
    \affiliation[b]{Kavli Institute for the Physics and Mathematics of the Universe (WPI), University of Tokyo, Chiba 277-8583, Japan}
\author[a]{Patrick Huber}
    \emailAdd{pahuber@vt.edu}
\author[a]{Ian M. Shoemaker\,\orcidlink{
0000-0001-5434-3744}}
    \emailAdd{shoemaker@vt.edu}

\abstract{Sterile neutrinos can be produced through mixing with active neutrinos in the hot, dense core of a core-collapse supernova (SN). The standard bounds on the active-sterile mixing ($\sin^2 \theta$) from SN arise from SN1987A energy-loss, requiring $E_{\text{loss}}<10^{52}~{\rm erg}$. In this work, we discuss a novel bound on sterile neutrino parameter space arising from the energy deposition through its decays inside the SN envelope. Using the observed underluminous SN IIP population, this energy deposition is constrained to be below $\sim 10^{50}~{\rm erg}$. Focusing on sterile neutrino mixing only with tau neutrino, for heavy sterile masses $m_s$ in the range $100$-$500$ MeV, we find stringent constraints on $\sin^2 \theta_\tau$ reaching two orders of magnitude lower than those from the SN1987A energy loss argument, {thereby probing the mixing angles required for Type-I seesaw mechanism}. Similar bounds will also be applicable to sterile mixing only with muons ($\sin^2 \theta_\mu$).}

\begin{document}
\maketitle
\flushbottom

\section{Introduction}
\label{sec:intro}
Although a great deal has been determined about neutrinos, the origin of neutrino mass remains unknown. A particularly simple possibility is that the Standard Model (SM) is augmented with at least two right-handed neutrinos which are singlets under SM interactions. As a result, such states can generate Dirac masses through nonzero Yukawa couplings to the SM lepton doublet and the Higgs, as well as Majorana masses at an unknown scale. After electroweak symmetry breaking, the left- and right-handed neutrinos mass mix, which effectively endows the ``sterile neutrinos'' with a suppressed coupling to the weak force, where the suppression is controlled by the mixing angles.  In addition to accounting for neutrino masses, sterile neutrinos may play a role in unravelling other mysteries as well, including acting as a dark matter candidate {(keV sterile)}~\cite{Dodelson:1993je,Shi:1998km}, and providing an origin for the observed baryon asymmetry~\cite{Akhmedov:1998qx}. 

In the face of significant theoretical uncertainty regarding the masses of sterile neutrinos, a wide range of varying experimental and theoretical probes have been brought to bear on the existence of sterile neutrinos over many orders of magnitude in mass~(e.g.,~\cite{Horiuchi:2013noa,Horiuchi:2015qri,Bolton:2019pcu,Abdullahi:2022jlv,Acero:2022wqg, Chauhan:2023pur}). Such searches include colliders, beta decays, accelerators, as well as astrophysical and cosmological signatures. 

In the present paper, we investigate the impact of 10-500 MeV scale sterile neutrinos on supernovae (SNe). Our work differs from previous studies on sterile neutrinos produced in SNe in several ways. In Refs.~\cite{Fuller:2008erj,Rembiasz:2018lok} it was found that $\sim$100 MeV scale sterile neutrinos can transport large quantities of energy and augment the explosion. Moreover strong limits on sterile neutrinos have been derived on the argument that excessive energy loss would catastrophically shorten the neutrino burst, in contradiction with observations; this has been applied to eV~\cite{Hidaka:2007se,Tamborra:2011is}, keV~\cite{Raffelt:2011nc,Suliga:2020vpz,Arguelles:2016uwb,Suliga:2019bsq,Syvolap:2019dat,Ray:2023gtu}, and $\sim 100$ MeV scale sterile neutrinos~\cite{Falk:1978kf,Dolgov:2000jw,Fuller:2008erj,Rembiasz:2018lok,Mastrototaro:2019vug}. 

Here however we derive strong constraints on sterile neutrinos based on the argument that they do not deposit more energy in the stellar envelope than what is observed in underluminous Type IIP SNe. Thus the lowest energy SNe yet observed will provide the strongest constraints. Similar arguments have been studied in axions \cite{Caputo:2022mah}.

\begin{figure}[htb!]
    \centering    \includegraphics[width=0.9 \textwidth]{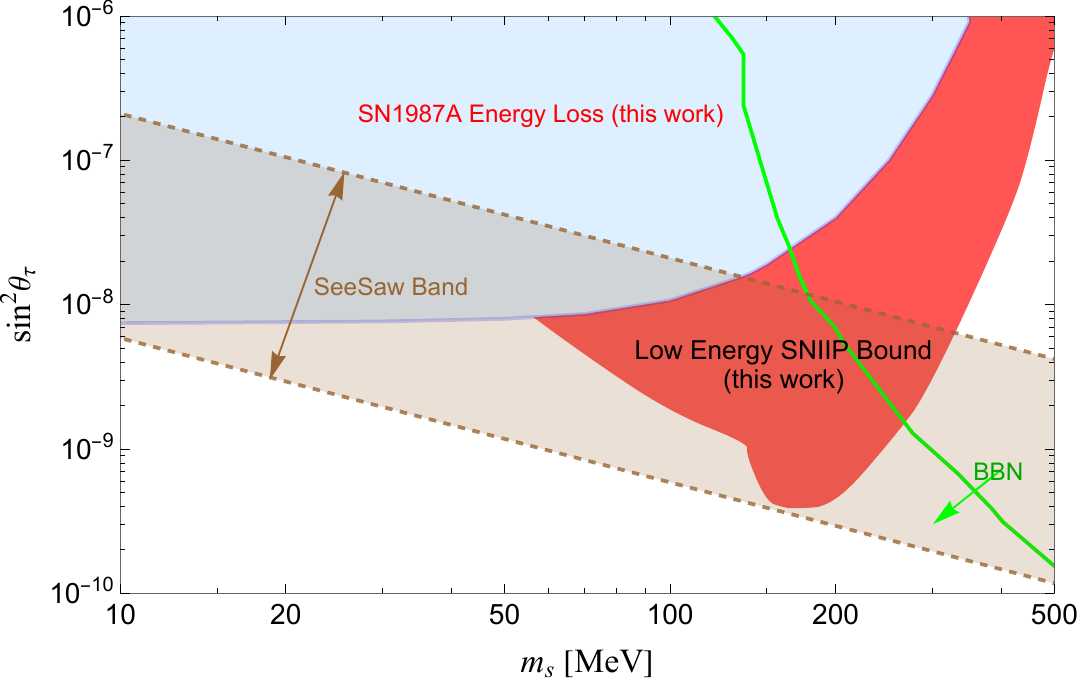}
    \caption{Bounds on sterile-active mixing $\sin^2 \theta_\tau$ as a function of sterile mass $m_s$.  The  \GC{red shaded region} shows the low-energy SN bound assuming full energy deposition from decaying sterile inside the SN envelope (\GC{Fe core $18.8 M_\odot$ progenitor}) \GC{extending upto $5\times 10^{12}$ cm}. The \GC{blue}-shaded region is the bound from energy loss in SN1987A calculated for the same $18.8 M_\odot$ progenitor. Other displayed constraints include BBN bounds assuming standard cosmology~\cite{Boyarsky:2009ix,Ruchayskiy:2012si,Sabti:2020yrt} \GC{and the parameter region favored by conventional Type-I seesaw models labeled `SeeSaw Band'}.}
    \label{fig:mainplot}
\end{figure}

\begin{figure}[ht!]
    \centering
    \includegraphics[width=0.49\textwidth]{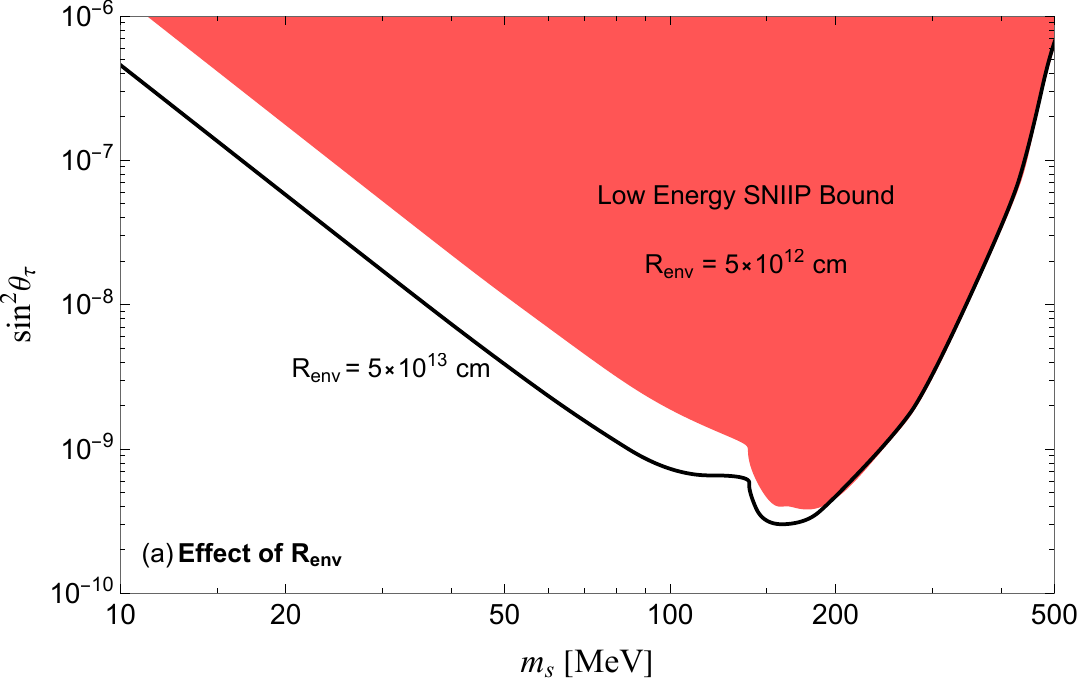}
    \includegraphics[width=0.49\textwidth]{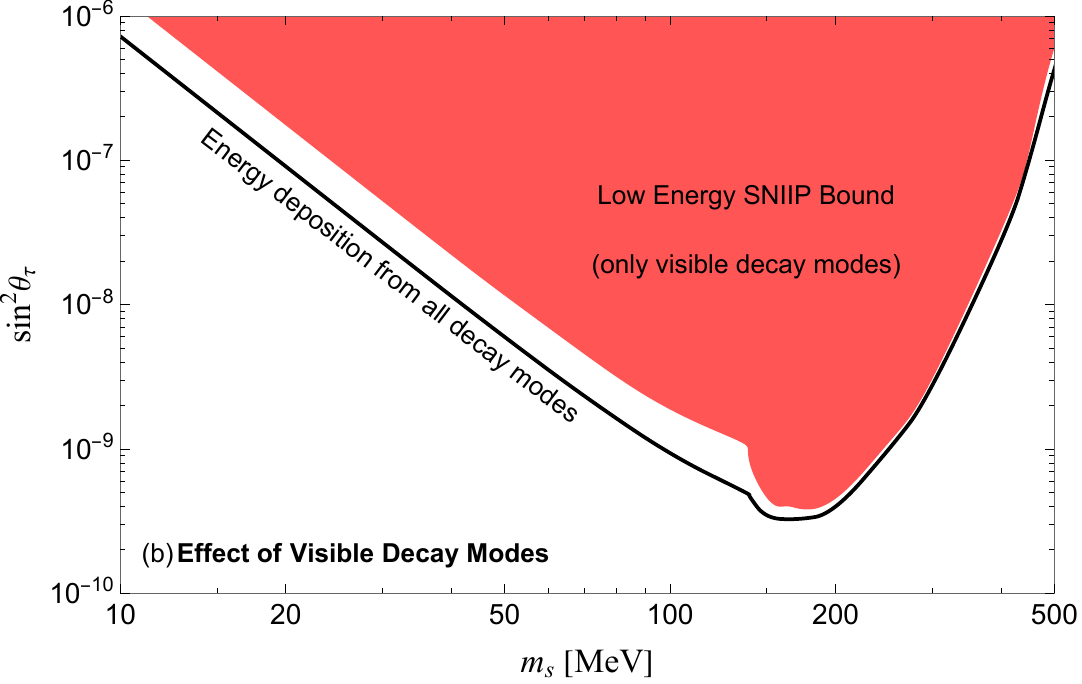}
    \includegraphics[width=0.49\textwidth]{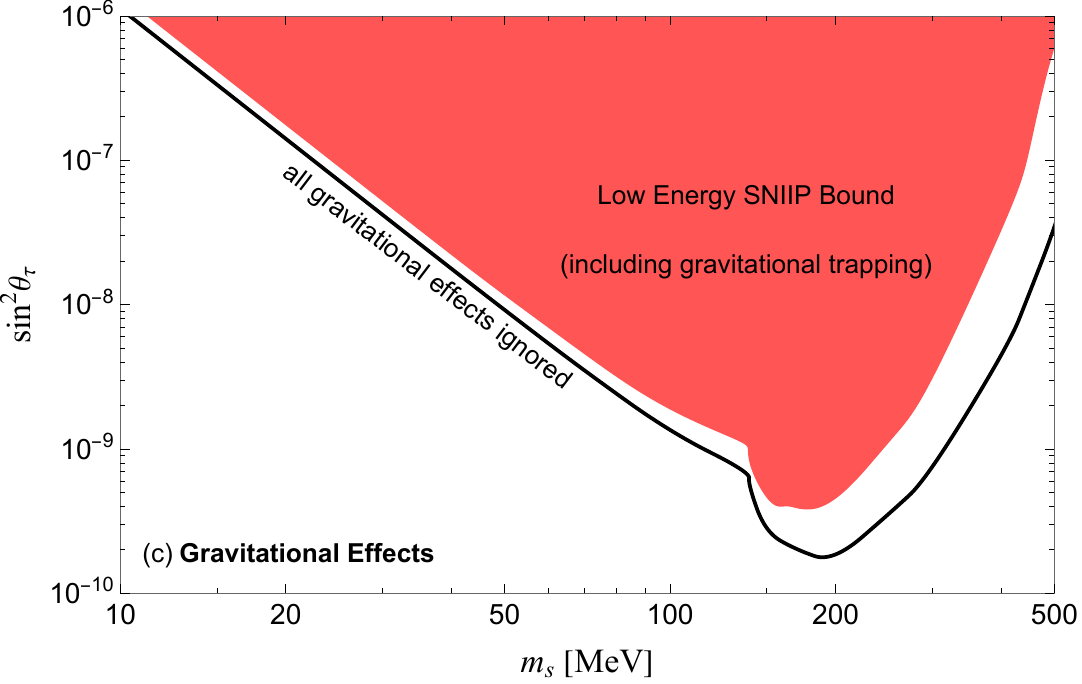}
    \includegraphics[width=0.49\textwidth]{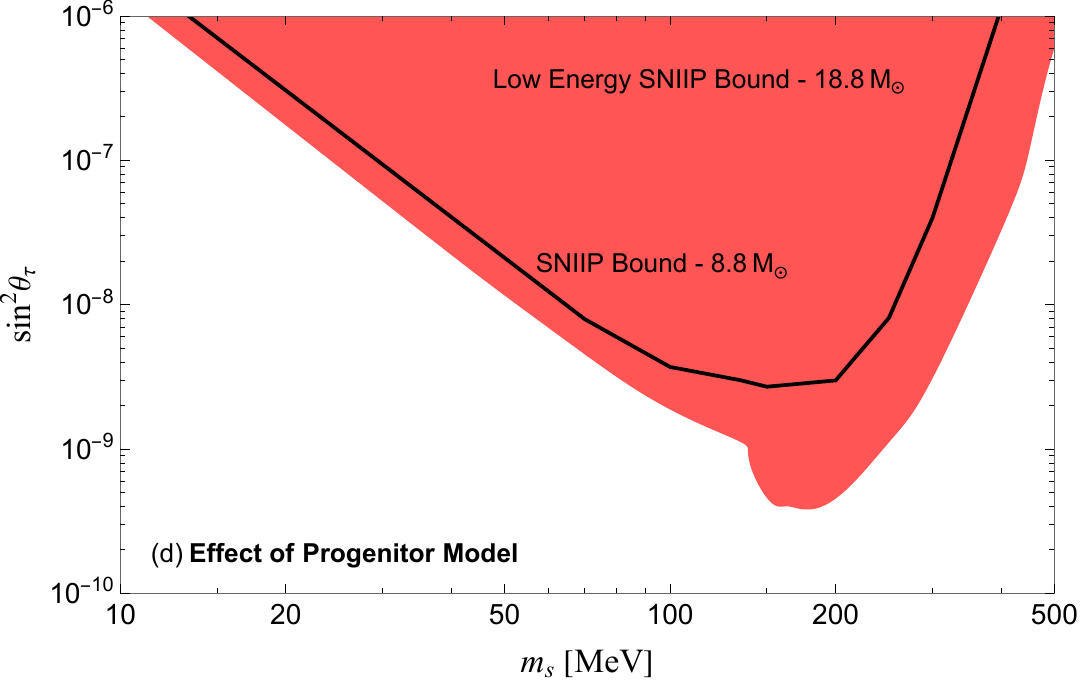}
    \caption{\GC{Comparison for bounds on sterile-active mixing $\sin^2 \theta_\tau$ as a function of sterile mass $m_s$, based on (a) Effect of the SN envelope size $R_{\text{env}}$, (b) Effect of energy deposition from visible decay modes vs. all decay modes, (c) Effect of gravitational trapping, (d) Effect of different progenitor models (shown for an Fe core $18.8 M_\odot$ progenitor and a ONeMg core $8.8 M_\odot$ progenitor, as labeled). The base model for constraint labeled `Low Energy SNIIP Bound' includes energy deposition from visible decay modes along with gravitational trapping effect taken into account for an Fe core $18.8 M_\odot$ progenitor with $R_{\text{env}}=5\times 10^{12}$ cm.} }
    \label{fig:modmainplot}
\end{figure}

As such, we extend the SM to include a heavy sterile neutrino $\nu_s$ which interacts with the SM exclusively via the type I seesaw Lagrangian~\cite{Minkowski:1977sc,Gell-Mann:1979vob,Mohapatra:1979ia},
\begin{equation}
\mathcal{L} \supset y_{\nu} \bar{L} \tilde{H} \nu_s + \frac{m_s}{2}\bar{\nu}^{c}_s\nu_s,
\label{eq:lagr}
\end{equation}
where $H$ and $L$ are the $SU(2)_{L}$ Higgs and lepton doublets respectively, $y_{\nu}$ is the neutrino Dirac Yukawa matrix, and $m_s$ is the Majorana mass term. \GC{After Higgs field develops a vacuum expecation value $v$, the resulting $(3+\nu_s)$ neutrino mass matrix can be block diagonalized by a matrix $W$ into light and heavy fields~\cite{Grimus:2000vj,Hettmansperger:2011bt},}
\begin{equation} 
W =
\begin{pmatrix}
\sqrt{1-BB^\dagger} & B \\
-B^\dagger &  \sqrt{1-B^\dagger B} 
\end{pmatrix}
\end{equation}
\GC{where $B^*\simeq m^\text{T}_D\, m_s^{-1}$ and $m_D=y_\nu\,v/\sqrt{2}$. In $(1+1)$ case, the $2\times 2$ orthogonal matrix can be parametrized simply by an active-sterile mixing angle $  B\simeq \sin \theta$.} 

At the low-scales relevant for our analysis, the main effect of Eq.~(\ref{eq:lagr}) is encoded in the active-sterile mixing angle. The parameter space SNe are most sensitive to, in the case of sterile-active mixing with electron ($\sin^2 \theta_e$) is already strongly constrained from the terrestrial probes. {For instance, the bounds on $\sin^2 \theta_e$ from terrestrial experiments especially for steriles heavier than $70$ MeV can already probe mixing angles upto $\sim 10^{-9}$~\cite{PIENU:2017wbj,T2K:2019jwa,NA62:2020mcv}}. In absence of muons in the SNe, the cases for sterile-active mixing with muon or tau neutrino are equivalent. Presence of muons can lead to extra channels for sterile neutrino production through charged-current interactions. Therefore in this work, we assume that the sterile neutrino mixes exclusively only with $\nu_\tau$. In terms of mass eigenstates $\nu_1$ and $\nu_2$, 
\begin{align}
     \nu_\tau & = \cos{\theta_\tau} \, \nu_1 + \sin{\theta_\tau} \, \nu_2 \nonumber \\ 
     \nu_s & = -\sin{\theta_\tau} \, \nu_1 + \cos{\theta_\tau} \, \nu_2      
\end{align}
where $\sin{\theta_\tau}$ denotes the mixing angle between the sterile and the tau neutrino.

\section{Low-Energy Supernovae}
\label{sec:leSN}
While SN1987A was unique in its proximity, SNe are not rare and can be used to study sterile neutrinos in many ways. We focus on a sub-class of core-collapse SNe with low-explosion energies, sometimes called underluminous Type IIP SNe, where the Type II represents the presence of hydrogen in the spectra and the P represents the SN's light-curve displaying a flat, or \textit{plateau}, shape in time. The luminosity and duration of the plateau reflects the explosion energy, ejecta mass, nickel ${}^{56}$Ni mass and progenitor radius. Therefore, the explosion energy can be inferred given the spectrum and the light curves. For example Refs.~\cite{Pejcha:2015pca,Muller:2017bdf,Goldberg:2019ktf,Murphy:2019eyu} used fitting formulae, simulations, and statistical inference along with quantifying the uncertainties, to infer the most likely explosion energies for a collection of Type IIP SNe. {The range of inferred mean explosion energies for the population spans $E_{obs}$ from $7.4 \times 10^{49}$ erg to $4 \times 10^{51}$ erg \cite{Murphy:2019eyu}}. While the upper end of the reconstructed energies are larger than typical predictions of numerical simulations, the lower end is approximately consistent with various simulated SNe. {It should be noted however that typical predictions have not been simulated to long enough times to observe the final saturated explosion energy, so predictions are in practice lower limits of explosion energy.}  {The spread in the reconstructed explosion energy for one of the lowest energy SNIIP candidate SN2001DC is $E_{\text{exp}}=10^{-1.13\pm 0.33} \times 10^{51}$ erg $\simeq 10^{49}-10^{50}$ erg~\cite{Murphy:2019eyu}.} Here, we are interested in sterile neutrinos produced in the core which can decay and deposit energies of a similar magnitude, and hence can be constrained from the {observations of lowest low-energy SNe, which we conservatively choose to be the $1\sigma$ upper bound on the explosion energy, i.e.,} $E_\text{dep}<10^{50}$ erg. Since numerical simulations are typically under-powered compared to observations, we expect our constraints to be conservative, unless some surprising systematic over-evaluation is discovered in simulations.   

We will examine two massive stars with initial masses $8.8 M_{\odot}$ and $18.8 M_{\odot}$ for obtaining bounds on the sterile neutrino parameter space. There are several reasons for these choices. First, the SN explosion energy is not a simple monotonic function of the initial mass of the progenitor. Instead, the explosion energy depends on the structure of the inner few solar masses of the progenitor's core, which in turn depends in a complex way to the progenitor mass, metallicity or even final hydrogen mass \cite{sukhbold:2014aa,Pejcha:2015pca}. Second, light progenitors have a systematically different core structure to those of Fe cored stars \cite{Jones:2013wda} which collapse by electron capture on its ONeMg core, which typically explode more readily with lower explosion energies \cite{Kitaura:2005bt,Hudepohl:2009tyy}. Our adopted $8.8 M_\odot$ progenitor aims to model this. Similarly, we adopt the $18.8 M_{\odot}$ as the typical massive star which evolves into a supergiant where its Fe core collapses causing a Type IIP SN. Finally, estimates of the progenitor masses of underluminous Type IIP supernovae cover a range: for example, SN1997D with estimated energy $1.0\times 10^{50}$~erg has an estimated initial mass $10\pm2 M_\odot$ \cite{Chugai:1999en}, while SN2003Z has estimated $1.6 \times 10^{50}$~erg and $14.15\pm 0.95 M_\odot$ \cite{Pumo:2016nsy} and SN2008kb $1.8 \times 10^{50}$~erg and $12.15\pm 0.75 M_\odot$ \cite{Pumo:2016nsy}. We thus consider the reality of the progenitors of underluminous Type IIP to lie somewhere between our adopted $8.8 M_{\odot}$ and $18.8 M_{\odot}$ progenitors. 

\subsection{Energy Deposition}
\label{sec:edep} 
Since the explosion energies of SN IIP are inferred from the light curves and dependent on the ejecta mass and the amount of $\prescript{56}{}{\text{Ni}}$ synthesized in the outer envelope, the limit on the energy deposition by any exotic species is constrained to the mantle region of the SN; in other words, we consider species which escape beyond the photosphere will simply present an energy sink for the explosion. The total energy deposited by sterile neutrino produced in the SN core, decaying outside the core but inside the SN envelope region ($R_{core}<r<R_{env}$) is given by 
\begin{align}
    E_{\text{dep}}  = \eta_\text{lapse}^2 & \int dt \int_0^{R_\text{core}} dr \int_{m_s}^{\infty} dE_s \frac{dL_s(r,E_s,t)}{dr\,dE_s}\Theta\left(E_s - \frac{m_s}{\eta_\text{lapse}}\right) \nonumber \\
    & \times \left\{ \text{exp}\left[-\frac{(R_{core}-r)}{L_{decay}}\right] - \text{exp}\left[-\frac{(R_{env}-r)}{L_{decay}}\right] \right\},
\end{align}
where  $\eta_\text{lapse}$ is the gravitational redshift factor, $E_s$ is the sterile neutrino energy, $\frac{dL_s(r,E_s,t)}{dr\,dE_s}$ is gradient of the differential sterile neutrino luminosity, $\Theta(x)$ is the Heaviside theta function and $L_{decay}$ is the decay length. Note that at large mixing angles, the decay length becomes comparable to the PNS radius and our treatment should not be taken as rigorous. {Although the decay length formalism does give very good estimate of the trapping regime but} we hope to return to this trapping regime in future work with greater precision. 

Sterile neutrinos produced in the SN core will also have to overcome the strong gravitational attraction arising from such high matter densities to avoid trapping. If sterile neutrino energy is sufficiently small, $E_s < m_s/\eta_\text{lapse}$, it will be gravitationally trapped inside the SN core.   The \textit{lapse} factor $\eta_\text{lapse}$ is the conversion factor relating the energy measured locally in the SN frame to the energy measured at the same point by an observer at infinity. For example in the weak-field limit of the Schwarzschild metric, the gravitational lapse factor is given by $\eta_\text{lapse}\simeq m_s(1+G\,M(r)/r)$. We also need to correct for the difference between the local time and the observer time in the energy emission rate. Both of these effects taken together lead to the inclusion of $\eta_\text{lapse}^2$ in the expression for $E_{\text{dep}}$. \GC{In our case, we find that our results are sensitive to the gravitational trapping (inside the core) only for heavy steriles}. Since energy spectrum for such heavy steriles is peaked at $E_N\sim M_N$, there is not enough kinetic energy available to escape the gravitational potential well and flux of the high energy tail of the spectrum that does escape, is not sufficient to deposit considerable energy greater than $10^{50}$ erg. \GC{Therefore, the SNIIP bound becomes less severe for such high masses.} 

\section{Sterile Neutrino Production}
\label{sec:nuSprod}
In a SN core, sterile neutrinos can be produced by $e^+$-$e^-$ or neutrino pair annihilation and the inelastic scattering of (anti-)neutrinos. \GC{In earlier literature ~\cite{Fuller:2008erj,Rembiasz:2018lok,Mastrototaro:2019vug,Syvolap:2023trc}, it was assumed that the strong degeneracy of $n,p,e^+,e^-,\nu_e,\Bar{\nu}_e$ in the hot proto-neutron star core will render pair-annihilation and inelastic scattering on the non-degenerate neutrino species ($\nu_\mu,\nu_\tau$) the dominant processes for $\nu_s$ production.} 

\GC{However in a recent work~\cite{Carenza:2023old}, the authors point out that the sterile neutrino production increases drastically by including the scattering of neutrinos off of neutrons and protons, i.e., $\nu+N \rightarrow \nu_s +N$, which had been missed by previous studies. We have verified their result by including this missing production channel and find that the $\nu_s$ production rate gets higher by 1-2 orders of magnitude.}

The amplitude for these relevant processes is given in Table.~\ref{table:sterileprod}. Note that we have explicitly listed the charge-conjugated processes in order to avoid any confusion. 
\begin{table}[h]
\centering
\begin{tabular}{|c c|} 
 \hline
 Process ($1+2\rightarrow 4+3$) & $S|M|^2$/($8 G_F^2 \sin^2\theta_\tau$) \\ [0.5ex] 
 \hline\hline
 $\nu_\tau+\Bar{\nu}_\tau\rightarrow \nu_s + \Bar{\nu}_\tau$ & $4u(u-m^2_s)$  \\ 
 \hline
 $\nu_\tau+\Bar{\nu}_\tau\rightarrow \nu_s + {\nu}_\tau$ & $4u(u-m^2_s)$  \\ 
 \hline
 $\nu_\mu+\Bar{\nu}_\mu\rightarrow \nu_s + \Bar{\nu}_\tau$ & $u(u-m^2_s)$  \\ 
 \hline
 $\nu_\mu+\Bar{\nu}_\mu\rightarrow \nu_s + {\nu}_\tau$ & $u(u-m^2_s)$  \\ 
 \hline
 $\nu_\tau+\Bar{\nu}_\tau\rightarrow \nu_s + {\nu}_\tau$ & $2s(s-m^2_s)$  \\ 
 \hline
 $\Bar{\nu}_\tau+\Bar{\nu}_\tau\rightarrow \nu_s + \Bar{\nu}_\tau$ & $2s(s-m^2_s)$  \\ 
 \hline
 $\nu_\mu+\Bar{\nu}_\tau\rightarrow \nu_s + {\nu}_\mu$ & $s(s-m^2_s)$  \\ 
 \hline
 $\Bar{\nu}_\mu+\Bar{\nu}_\tau\rightarrow \nu_s + \Bar{\nu}_\mu$ & $s(s-m^2_s)$  \\ 
 \hline
 $\nu_\tau+\Bar{\nu}_\mu\rightarrow \nu_s + \Bar{\nu}_\mu$ & $u(u-m^2_s)$  \\ 
 \hline
 $\nu_\mu+\Bar{\nu}_\tau\rightarrow \nu_s + {\nu}_\mu$ & $u(u-m^2_s)$  \\ 
 \hline
 $\nu_\tau(\bar{\nu}_\tau)+ e^{-}\rightarrow \nu_s + e^{-}$ & $8 [g_L^2(p_1\cdot p_2)(p_3\cdot p_4) + g_R^2(p_1\cdot p_3)(p_2\cdot p_4)-g_Lg_Rm_e^2(p_1\cdot p_4)]$  \\ 
 \hline
 $\nu_\tau(\bar{\nu}_\tau)+ e^{+}\rightarrow \nu_s + e^{+}$ & $8 [g_R^2(p_1\cdot p_2)(p_3\cdot p_4) + g_L^2(p_1\cdot p_3)(p_2\cdot p_4)-g_Lg_Rm_e^2(p_1\cdot p_4)]$  \\ 
 \hline
  $e^{-}+ e^{+}\rightarrow \nu_s + \nu_\tau(\bar{\nu}_\tau)$ & $8 [g_R^2(p_1\cdot p_4)(p_2\cdot p_3) + g_L^2(p_1\cdot p_3)(p_2\cdot p_4)-g_Lg_Rm_e^2(p_3\cdot p_4)]$  \\ 
 \hline
 $\nu_\tau(\bar{\nu}_\tau)+ N\rightarrow \nu_s + N$ & $|\mathcal{M}|^2_{VV}+|\mathcal{M}|^2_{AA}+|\mathcal{M}|^2_{VA}$  \\ 
 \hline
\end{tabular}
\caption{Matrix element squared $S|M|^2$ for the dominant processes involved in sterile neutrino production in units of $8 G_F^2 \sin^2\theta_\tau$~\cite{Fuller:2008erj,Carenza:2023old} \GC{, where $g_L=-0.5+\sin^2{\theta_W}$ and $g_R=\sin^2{\theta_W}$. The complete expression for neutrino-nucleon scattering is provided in Appendix.~\ref{app:A}.}}
\label{table:sterileprod}
\end{table}

\subsection{Boltzmann Transport}
\label{sec:BoltzT}
\begin{figure}[t!]
    \centering
    \includegraphics[width=0.7\textwidth]{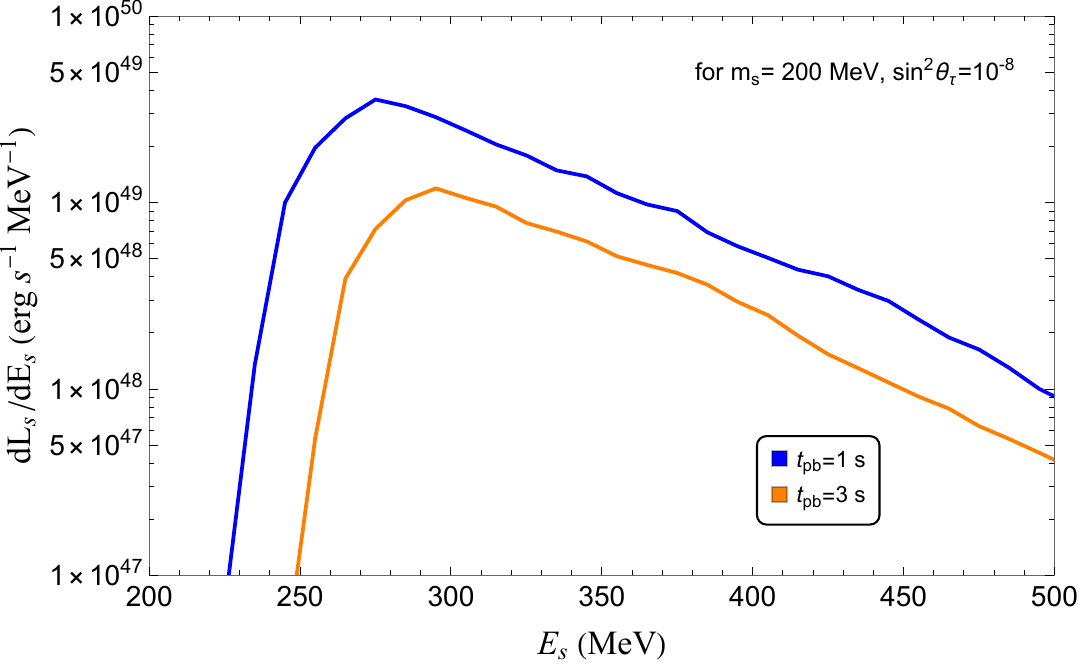}
    \caption{\GC{Differential sterile neutrino luminosity for sterile neutrino mass $m_s= 200$ MeV and $\sin^2{\theta_\tau} = 10^{-8}$ at different post-bounce times, see e.g., Eq.~(\ref{eq:diff}).}}
    \label{fig:diffLum}
\end{figure}

\begin{figure}[tb!]
    \centering
    \includegraphics[width=0.7\textwidth]{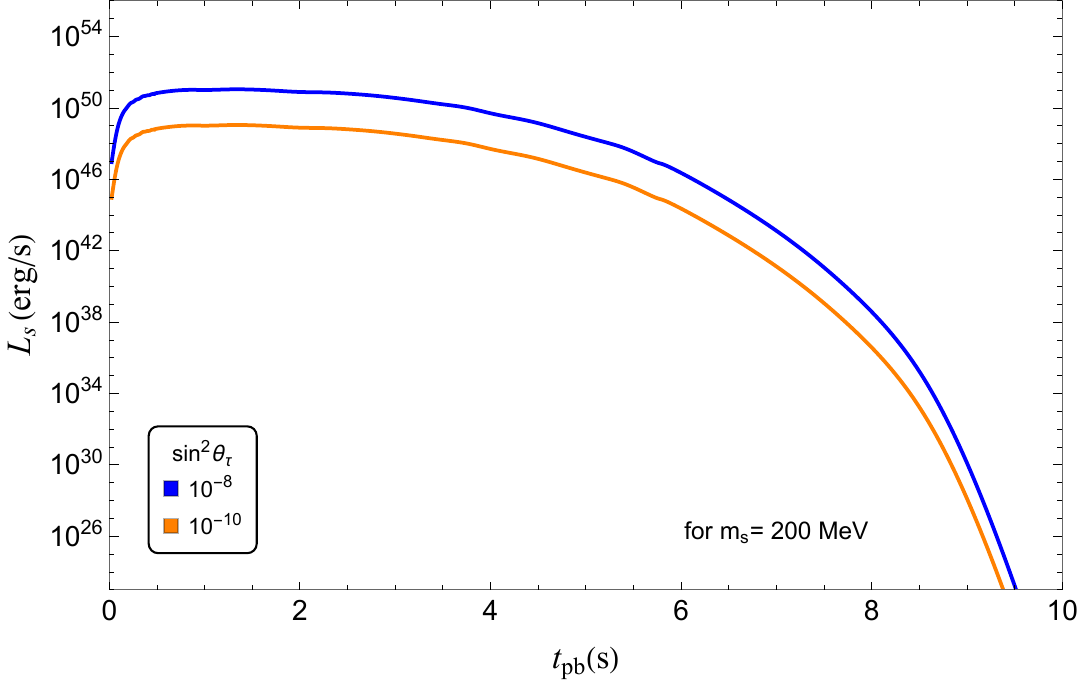}
    \caption{Integrated sterile neutrino luminosity plotted as a function of post-bounce time for sterile neutrino mass $m_s= 200$ MeV for different $\sin^2{\theta_\tau}$, see Eq.~(\ref{eq:diff}).}
    \label{fig:intLum}
\end{figure}

The evolution of sterile neutrino abundances is governed by the Boltzmann transport equation. Since solving the exact transport equation {is beyond the scope of the current study}, we simplify the task at hand {for our purpose} by assuming the medium is homogeneous and isotropic \cite{Hannestad:1995rs}. This implies that the change in phase-space density will only be affected by the scatterings/pair-annihilation processes in the SN core. In this case, the simplified kinetic equation for sterile neutrino production is 
\begin{equation}
    \frac{\partial f_s}{\partial t} = \mathcal{C}_{coll}(f_s),
\end{equation}
where $f_s$ is the sterile neutrino phase-space density distribution and $\mathcal{C}_{coll}$ is the sum of all possible collisional interactions.
In our case, the collisional term for $2\rightarrow 2$ particle interactions can be written as
\begin{align}
    \mathcal{C}_{coll} = &\frac{1}{2 E_s} \int d^3 \Tilde{p_2} d^3 \Tilde{p_3} d^3 \Tilde{p_4}\, \Lambda(f_s,f_2,f_3,f_4) \times \nonumber \\ &  S|M|^2_{12\rightarrow 34}\, \delta^4(p_s+p_2-p_3-p_4) (2 \pi)^4,
\end{align}
where $d^3 \Tilde{p_i}= d^3 {p_i}/((2 \pi^3)\,2 E_i)$, $\Lambda(f_s,f_2,f_3,f_4)= (1-f_s)(1-f_2)f_3 f_4- f_sf_2(1-f_3)(1-f_4)$ is the phase-space factor including the Pauli blocking of the final states, $S$ is the symmetry factor, $|M|^2$ is the interaction matrix element element squared, $E_i$ and $p_i$ are energy and momentum of the $i$-th particle with subscript label $s$ for sterile neutrino. For the range of interest for $\sin^2 \theta_\tau$, the sterile neutrino produced will not be trapped in the SN, hence we can safely assume $f_s=0$. After numerically solving for $f_s$, we can calculate the sterile neutrino luminosity $\frac{dL_s}{dE_s}$ as~\cite{Tamborra:2017ubu,Mastrototaro:2019vug}
\begin{equation}
    \frac{dL_s}{dE_s} = \frac{2 E_s}{\pi} \int dr\,r^2 \frac{df_s}{dt} E_s\,p_s.
    \label{eq:diff} 
    \end{equation}
As an example, we show this differential and integrated sterile neutrino luminosity in Figs.~\ref{fig:diffLum} and~\ref{fig:intLum}, respectively for reference sterile neutrino mass $m_s= 200$ MeV. 

\subsection{Visible Energy Deposition}
\label{sec:visEdep}
The sterile neutrino decays and deposits energy into the SN envelope. For the mass range of interest and mixing only with $\nu_\tau$, the charged-current processes are kinematically forbidden. Therefore, we only consider following neutral current decays and their charge-conjugate processes, assuming $\nu_s$ to be Majorana particles ~\cite{Gorbunov:2007ak, Atre:2009rg, Bondarenko:2018ptm,Coloma:2020lgy,Helo:2010cw}. Note that the analytical expressions for $\pi^0$ decay mode width in Refs.~\cite{Gorbunov:2007ak,  Bondarenko:2018ptm,Coloma:2020lgy} differ by a factor of $2$ (in the numerator) compared to Refs~\cite{ Atre:2009rg,Helo:2010cw}. In this work, we use expressions given in Ref.~\cite{Coloma:2020lgy}. The relevant decay modes for $\nu_s$ are
\begin{align}
    \nu_s &\rightarrow \nu_\tau \pi^0, \nonumber \\
    \nu_s &\rightarrow \nu_\tau e^+e^-, \nonumber \\
    \nu_s &\rightarrow \nu_\tau \mu^+\mu^- ,\nonumber \\
    \nu_s &\rightarrow \nu_\tau\nu_x\Bar{\nu}_x,
\end{align}
where $x=(e,\mu,\tau)$. The process involving neutral pion tends to be the dominant decay mode for higher masses while the electron channel contributes negligibly to the decay width. We classify the first 3 listed processes as the \textit{visible} decay modes while the last decay mode entirely into neutrinos as \textit{invisible}. Based on this classification, we will discuss two limiting cases of energy deposition, $i$) the entire energy from sterile decay is deposited, and $ii$) only the \textit{visible} decay modes deposit their energy into the SN envelope. The energy deposition in the latter case is determined by the branching ratio into the \textit{visible} modes, given by
\begin{equation}
    BR_{vis} = 1-\frac{\Gamma(\nu_s \rightarrow \nu_\tau\nu_x\Bar{\nu}_x)}{\Gamma_{tot}},
\end{equation}
where $\Gamma_{tot}$ is the total decay width calculated from all four decay modes. 

We now discuss the efficacy of the branching ratio method. Two possible issues can be raised for this approach. Firstly, the presence of a neutrino final state in the the first 3 decay modes might take away a portion of the (assumed to-be) deposited energy. Secondly, there might be possible energy deposition into the mantle from the secondary neutrinos produced in the invisible decay channel. Before we address both of these issues, a crucial fact to remember is that neutrinos can get trapped or multiply scatter in the hot and highly dense SN environment, conditions for which can mainly occur in/near the proto-neutron star core. In the first case for $m_s \geq m_{\pi}$, the decays of the sterile neutrino are dominated by the pion mode. For mixing angles of interest, such steriles decay far inside the $R_{env}$, which gives $\nu_\tau$ produced to encounter high densities and undergo at least one scattering to deposit substantial energy into the envelope\footnote{Note that our discussion only centers around the decay lengths for the average secondary neutrino energies in a distribution. Some of these neutrinos if not being trapped/scatter, doesn't imply all of the produced neutrino in the decay mode under consideration will free stream. The higher end tail of the energy spectrum will most likely be deposited.}. For $m_s \leq m_{\pi}$, the decays are dominated by the 3$\nu$ mode, with average neutrino energy $\sim 80$ MeV. Such energetic neutrinos would have been trapped if produced near the core but for lighter $\nu_s$ (and relevant mixings), the average decay length is far outside the SN. Hence, a major portion of the neutrino energy will not be deposited and hence considering this channel as ``invisible'' is a good approximation. From the above discussion, we can conclude that the branching ratio method is a good estimator of the actual energy deposition from sterile neutrino decays. A proper inclusion of secondary neutrino energy deposition is beyond the scope of the current work and we stress that it should not affect very strongly the results presented here.

\section{Reference SN Models}
\label{sec:refSN}
In this work, we adopt two different SN progenitors to obtain the bounds in the mixing-mass plane as well as to test the robustness of the bounds to the SN progenitor used. Since we are concerned with very small mixing angles, we assume that the $\nu_s$ production does not affect the standard SN processes. In the \textit{Garching model}, we apply our reasoning to obtain realistic bounds with the SFHo-18.8 model simulated by the Garching group, which adopts a $18.8\, M_\odot$ progenitor and includes six-species neutrino transport \cite{ SNprofile,Mirizzi:2015eza,Bollig:2020xdr}. The SFHo EoS \cite{Steiner:2012rk} is used and PNS convection is modeled by a mixing-length treatment \cite{Bollig:2017lki}. We use the simulated SN evolution assuming $R_\text{core}\sim 20$ km for all post-bounce time sequences up to $\sim 10$s and envelope extending up to $\sim 5 \times 10^{7}$ km. 

We also consider a $8.8\, M_\odot$  SN progenitor \cite{Hudepohl:2009tyy} (which is on the lower end of the progenitor mass range that finally end up in a neutron star) to study the effect of progenitor dependence on our bounds. This progenitor collapses via electron-capture on its ONeMg core with a final baryonic mass of $1.366 \, M_\odot$ using Shen's stiff baryonic equation of state~\cite{Shen:1998aa} for hot dense nuclear matter and a final neutron star radius of about $15$ km. Also note that unlike the $18.8\, M_\odot$ progenitor model, this simulated progenitor does not include muons or convection in proto-neutron star. Although both of these models have peak temperatures of about $30-40$ MeV, we find quite different bounds in each case \GC{(See lower right panel in Fig.~\ref{fig:modmainplot} and discussion on results)}. 

\begin{figure}[t!]
    \centering
    \begin{subfigure}
        \centering
        \includegraphics[width=0.49\textwidth]{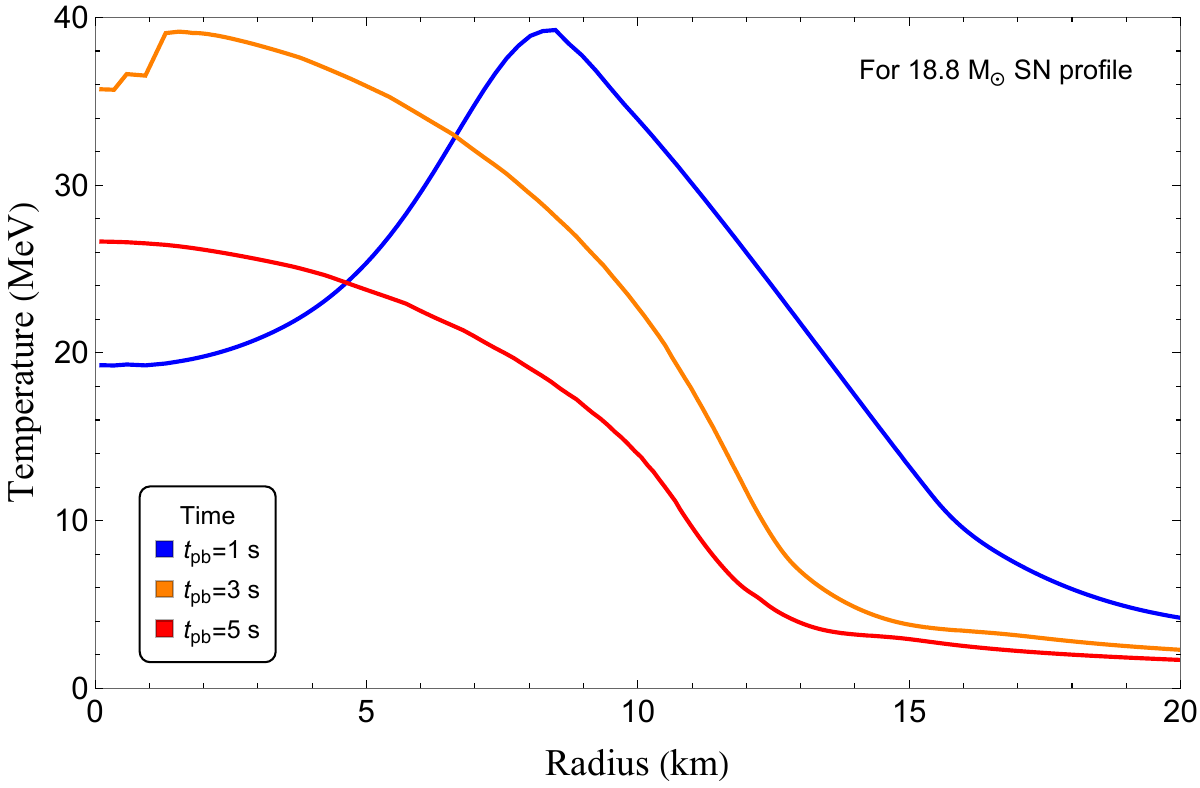}
    \end{subfigure}%
    \begin{subfigure}
        \centering
        \includegraphics[width=0.49\textwidth]{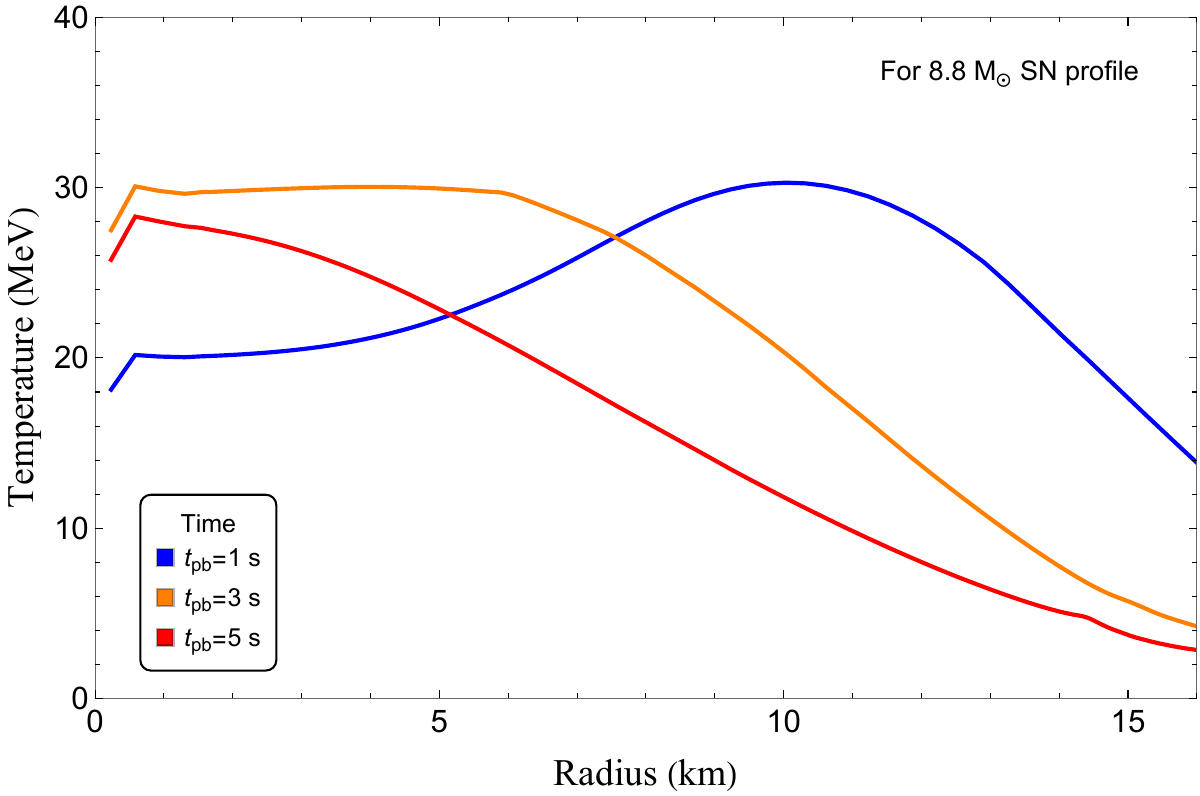}
    \end{subfigure}%
    \begin{subfigure}
        \centering
        \includegraphics[width=0.49\textwidth]{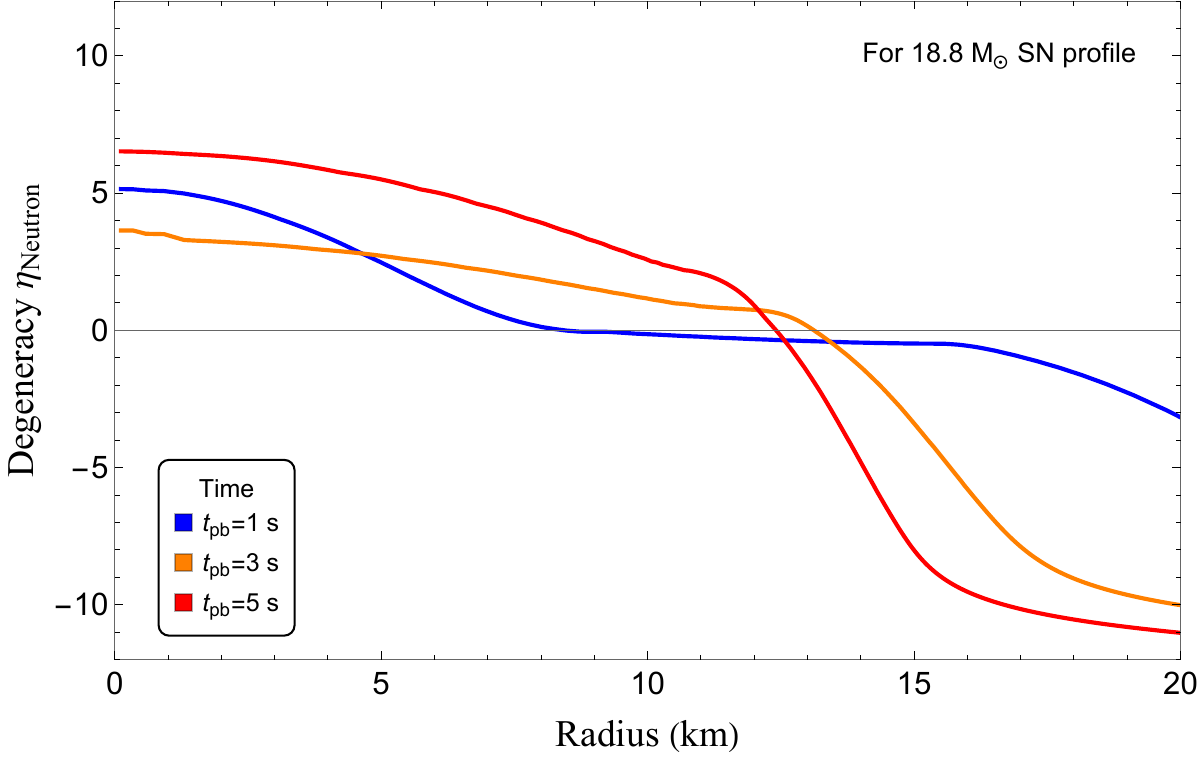}
    \end{subfigure}
    \begin{subfigure}
        \centering
        \includegraphics[width=0.49\textwidth]{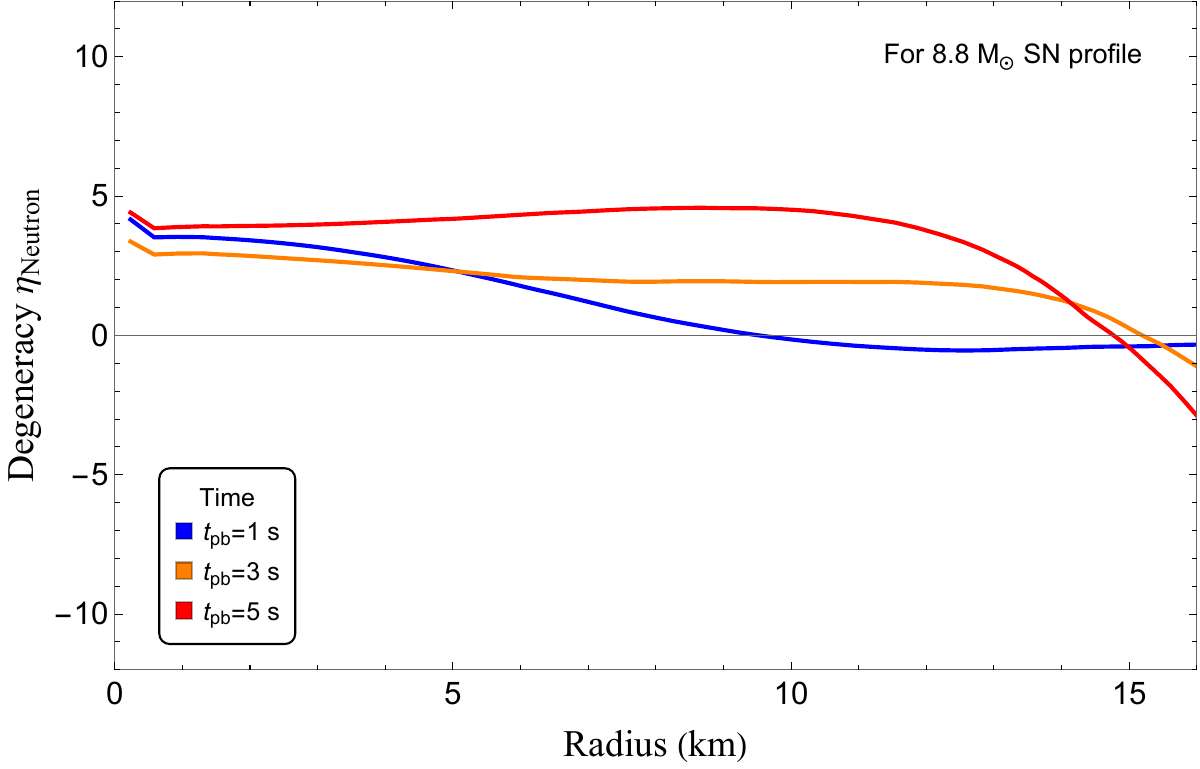}
    \end{subfigure}
    \caption{Temperature (upper panels) and \GC{$\eta_N$ neutron degeneracy parameter profiles} (lower panels) for 18.8 $M_{\odot}$ and 8.8 $M_{\odot}$ SN progenitors used in this work, at different post-bounce times $t_{\text{pb}}=(1,3,5)$ s.}
    \label{fig:tempchem}
\end{figure}
{The SNIIP explosion bound discussed in this work has been calculated for two different SN progenitors : (i) 18.8 $M_{\odot}$ (ii) 8.8 $M_{\odot}$. We provide time snapshots of the radial profiles depicting variations in the temperature and magnitude of the chemical potential of the muon neutrino in both of these progenitors, shown in Fig~\ref{fig:tempchem} at different post-bounce times $t_{\text{pb}}=$ 1,3 and 5 seconds.}   

{It can be clearly seen that average temperatures in 18.8 $M_{\odot}$ progenitor are higher (up to $\sim 40$ MeV) than 8.8 $M_{\odot}$ (with highest temperatures around 30 MeV). The sterile neutrino production rates is also dependent on muon neutrino chemical potential, which again are comparatively higher for 18.8 $M_{\odot}$ progenitor than the 8.8 $M_{\odot}$ progenitor. We note two other important points that the $\nu_\tau$ has vanishing chemical potential in both cases and the production rate directly doesn't depends on the density.}

The sterile neutrino production rate depends mainly on the sterile energy, sterile mass and the radial distribution of temperature. For SFHo-18.8, the outer layer of the core is the hottest in the first few time steps, as the in-falling matter towards the iron core first heats up the core edge. Thus, the maximum production occurs at edges of the core for the first few ms after $t_{pb}=0$. But as the inner core gets hotter with time, the maximum production rate shifts closer to the center. The total production rate grows with time up to $\sim 3\mbox{ -- }4$ s after which it falls off rapidly. 

It is easy to see that for masses $m_s \lesssim T$, the high energy sterile states can be produced easily. For heavier steriles, although mainly produced at rest, energies up to $\sim (6\mbox{ -- }8)\, T$ are feasible (since production is through $2\rightarrow2$ scattering processes). However, some higher energy states for heavier steriles can still be produced with modest Boltzmann suppression. 

\begin{figure}[t!]
    \centering
    \begin{subfigure}
        \centering
        \includegraphics[width=0.49\textwidth]{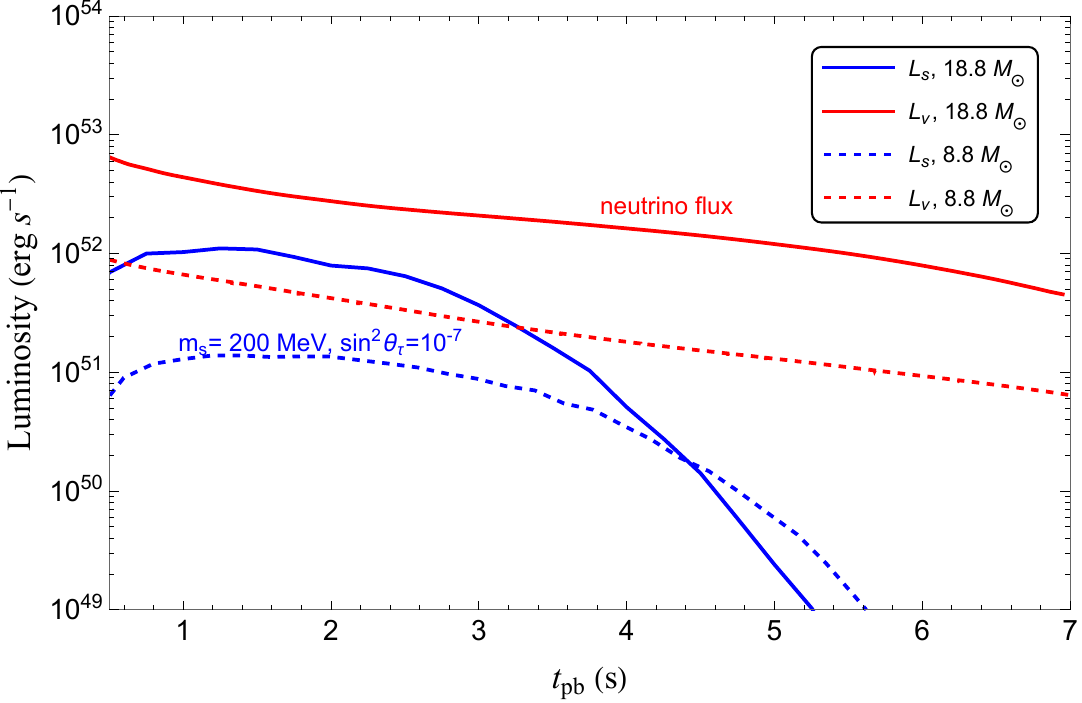}
    \end{subfigure}%
    \begin{subfigure}
        \centering
        \includegraphics[width=0.49\textwidth]{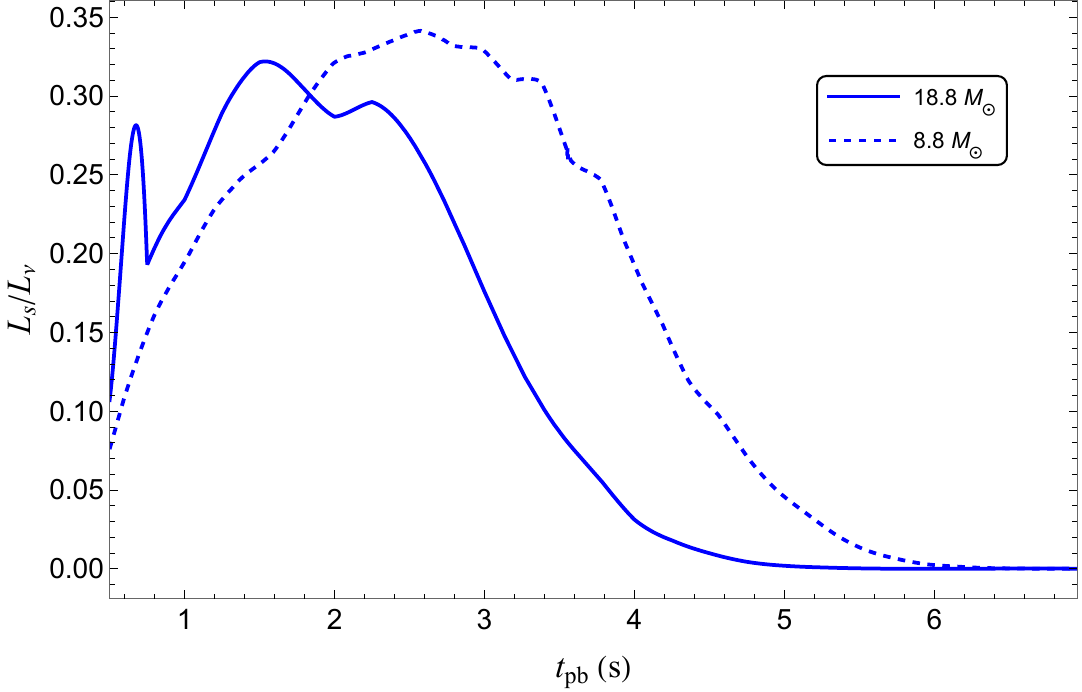}
    \end{subfigure}%
    \caption{(Left) Luminosity lightcurves as function of post-bounce time ($t_{\text{pb}}$) for active neutrinos (red) and sterile neutrinos (blue) for the 18.8 $M_{\odot}$ (solid) and 8.8 $M_{\odot}$ (dashed) SN progenitors considered in this work. Sterile neutrino parameters are set to $m_s=200$ MeV and $\sin^2{\theta_\tau}=10^{-7}$. (Right) The ratio of sterile neutrino to active neutrino luminosity for the same set of parameters as shown in the left panel.}
    \label{fig:nulum}
\end{figure}
{For better comparison of the dependence on the SN progenitor, we also plot the luminosity lightcurves for active neutrinos and sterile neutrinos as functions of the post-bounce time ($t_{\text{pb}}$), for both 18.8 $M_{\odot}$ and 8.8 $M_{\odot}$ SN progenitors in Fig~\ref{fig:nulum} (left panel). The sterile neutrino mass and mixing angles have been set to $m_s=200$ MeV and $\sin^2{\theta_\tau}=10^{-7}$ respectively. It can be seen that the sterile neutrino luminosity is subdominant compared to the active neutrino luminosity in both progenitors. However, the net fluxes for both sterile neutrino as well as active neutrino fluxes differ by approximately an order of magnitude. In addition, we also plot the ratio of sterile neutrino to active neutrino luminosities in the right panel. Note that the ratio should always be less than 1 for a consistent treatment of sterile neutrino production, which is indeed the case for $m_s=200$ MeV and $\sin^2{\theta_\tau}=10^{-7}$.}

\subsection{Fitting Functions}
Ideally, scaling behaviors can be extracted exactly for the sterile neutrino production rate. However, such scaling behaviors should not be expected to hold exactly in the case of space and time dependent quantities. The temperature and chemical potentials that determine the luminosity vary significantly depending on the radial position and the post-bounce time for the simulated SN progenitors. Thus, the total luminosity for each SN progenitor will be expected to show different scaling behavior with respect to temperature, chemical potential and masses.

For simplicity, we have extracted approximate fitting functions for the sterile neutrino luminosity for both progenitors from the production rates calculated in this work. \GC{The approximate fitting functions valid for $m_s>10$ MeV are 
\begin{equation}
L_s^{18.8\,M_{\odot}} = 1.161 \times 10^{53} \,\frac{\text{erg}}{\text{s}}\left( \frac{\sin^2 \theta_\tau}{10^{-7}}\right) \left( \frac{T}{40\, \text{MeV}}\right)^{7.55} \left( \frac{|\eta_N^{\text{min}}|}{1.0}\right)^{1.52} \text{Exp}\left(\frac{-m_s}{T}\right)
\label{eq:fit1}
\end{equation}
\begin{equation}
L_s^{8.8\,M_{\odot}} = 4.88 \times 10^{53} \,\frac{\text{erg}}{\text{s}}\left( \frac{\sin^2 \theta_\tau}{10^{-7}}\right) \left( \frac{T}{30\, \text{MeV}}\right)^{6} \left( \frac{|\eta_N^{\text{min}}|}{0.1}\right)^{-0.8} \text{Exp}\left(\frac{-m_s}{T}\right)
\label{eq:fit2}
\end{equation}}
\GC{These luminosity fitting functions should be evaluated with temperature $T$ and degeneracy parameter $|\eta_N|$ ($\equiv (\mu_N^*-m_N^*)/T$) corresponding roughly to the maximum and minimum values of these quantities respectively, for a given post-bounce time.} As can be seen in both cases, $L_s$ strongly depends on the temperature along with a exponential Boltzmann suppression factor. Note that these fitting functions are applicable in the regime where the mean free path of sterile neutrino exceeds $R_{\text{core}}$. 

Using these fitting functions, we can estimate the relative ratio of the bounds expected from two different progenitors. \GC{The integrated luminosity from the fitting function for $m_s=200$ MeV and $\sin^2{\theta_\tau}=10^{-7}$, $E^{8.8\,M_{\odot}}_{\text{int}}$ is $\sim 3.5 \times 10^{51}$ erg/s while for $E^{18.8\,M_{\odot}}_{\text{int}}$ is $\sim 2.5 \times 10^{52}$ erg/s, also in accordance with the sterile neutrino flux shown in fig.~\ref{fig:nulum} (left panel). The duration of dominant production lasts about $3$ s for both 18.8 $M_{\odot}$ and 18.8 $M_{\odot}$ SN progenitors. Since for such heavy mass and strong coupling, the sterile neutrino can decay inside the SN envelope, the total deposited energy in each case is $0.84\,E_{\text{int}}$, where the factor of $0.84$ is the fraction of sterile neutrino decays into {\it{visible}} decay modes for $m_s=200$}. On comparing these deposited energies, we would expect the SNIIP bound at $m_s=200$ MeV for 18.8 $M_{\odot}$ to be an order of magnitude stronger than 8.8 $M_{\odot}$ SN progenitor. This is in accordance with results obtained in Fig.~\ref{fig:modmainplot}. In principle, given the respective SN progenitors, Eqs.~\eqref{eq:fit1} and ~\eqref{eq:fit2} can be roughly used to obtain bounds in Fig.~\ref{fig:mainplot}.  

\section{Results}
\label{sec:results}
\GC{We display our main results in Fig.~\ref{fig:mainplot} only assuming non-zero tau neutrino mixing with the sterile state. For the red shaded region, we assume no more than $10^{50}$erg energy deposition for the 18.8 $M_{\odot}$ progenitor, and only include contributions from the electromagnetic decay products of the sterile. We also calculate the SN energy loss bound for the  18.8 $M_{\odot}$ progenitor, shown as light blue shaded region. We see that the low-energy SN constraint can be close to two orders of magnitude stronger than the energy loss bounds obtained from SN1987A~\cite{Mastrototaro:2019vug}, and provides the leading constraint on the mixing angle for $60~{\rm MeV} \lesssim m_{s} \lesssim 500$ MeV. Also included in Fig.~\ref{fig:mainplot} is the region favored by the canonical Type-I seesaw models for neutrino masses (dashed brown curves) and BBN bounds assuming standard cosmology~\cite{Boyarsky:2009ix,Ruchayskiy:2012si,Sabti:2020yrt,Domcke:2020ety}. We note that the BBN bounds exclude the region below the green curve in Fig.~\ref{fig:mainplot}. However, in the presence of a lepton asymmetry these constraints can be substantially weakened~\cite{Gelmini:2020ekg}.}

\GC{We also discuss the effects of SN envelope size, gravitational trapping, energy deposition and different progenitor models in Fig.~\ref{fig:modmainplot}. The base model for constraint labeled `Low Energy SNIIP Bound' is shown as red shaded region and includes energy deposition from visible decay modes along with gravitational trapping effect taken into account for an Fe core $18.8 M_\odot$ progenitor with $R_{\text{env}}=5\times 10^{12}$ cm. In the first panel, it can be clearly seen that a larger SN envelope size provides a larger volume for energy deposition by sterile neutrino decays, therefore leading to a stronger constraint. The enhancement is quite significant for lighter sterile neutrino masses, since their high boost factor compared to heavier steriles require additional time for same energy deposition through decays. From the upper right panel, we find that the constraint could be similarly stronger if all the sterile decay products including the active neutrinos deposit energy in the envelope. By including the effect of gravitational trapping (lower left panel), the constraint becomes weaker for heavier sterile masses especially for $m_s>100$ MeV. Finally, for the lower right panel, we see that the bound is significantly weakened if the progenitor mass is lowered to 8.8 $M_{\odot}$. We reason that our bounds depend strongly on the progenitor model is due to the fact that sterile neutrino production depends sensitively on $T$. The temperature profiles for $18.8 \,M_{\odot}$ SN have consistently higher temperatures for all $t=(0-10)$s as compared to the $8.8 \,M_{\odot}$ progenitor. It should be noted however that even in this case, our results provide constraints stronger than the cooling bound in the 100-400 MeV mass window.}

We note that while we have fixed to mass-mixing with the tau-neutrino in Fig.~\ref{fig:mainplot}, we expect qualitatively similar results for mixing with the muon-neutrino.

\section{Conclusion and Future directions}
\label{sec:future}
The present paper has examined the implications of heavy sterile neutrinos for low-energy Type IIP SNe (SNIIP). We pointed out that sterile neutrino decays can over-power SNe beyond the energies of SNIIP, thus placing stringent limits on sterile neutrino parameter space that are beyond those from SN1987A and potentially reach into the regime of Type-I seesaw models. 

This work can be extended in a number of directions. {Firstly, in our work, we did not perform a dedicated numerical treatment of core collapse and the impacts of in-situ sterile neutrino production and decay. Instead, the main aim of our work was to motivate that the bound from SNIIP explosion energy is substantially better than existing limits from SN1987A cooling bound by $1-2$ orders of magnitude. Thus, our results motivate  future detailed numerical investigations of heavy sterile neutrinos in core-collapse supernovae}. {Secondly}, we have only considered mixing with one active neutrino flavor at a time. It would be worthwhile to examine more generic flavor structure assumptions. {Thirdly}, it is possible that the production and decay of sterile neutrinos is not controlled by the EW force. For example, sterile transition magnetic moments have been widely studied~\cite{Gninenko:1998nn,Coloma:2017ppo,Magill:2018jla,Brdar:2023tmi}. We also anticipate low-energy SNe placing stringent constraints on transition magnetic moments~\cite{Magill:2018jla,Brdar:2023tmi,Chauhan:2024nfa}.   

Moreover, even while restricting to mass-mixing, there are additional avenues to be explored. Given that SNe are efficient production sites of sterile neutrinos, the impact of the escaping sterile neutrinos (and their decay products) could be significant. In particular, for axions it is known that requiring that the axions not produce a photon flux above the diffuse gamma-ray background leads to strong bounds at low masses~\cite{Caputo:2021rux,Diamond:2023scc}. Secondly, the high-energy neutrino flux produced by the decaying sterile neutrinos may also be detectable at future neutrino experiments like DUNE and Hyper-K~\cite{Mastrototaro:2019vug}, and a detailed follow-up in light of our results may yield new constraints. 

\GC{In conclusion, we have found that low-energy SNe can provide stronger constraints than SN1987A energy-loss bound on sterile neutrinos in the 60-500 MeV range.} In addition, these bounds probe theoretically well-motivated parameter space predicted by Type-I seesaw models of neutrino masses. In future work we plan to return to this scenario and explore implications of the neutrino and gamma-rays emerging from sterile decays beyond the SN envelope. 

\vspace{0.1cm}
\section*{Acknowledgements}
We are very grateful to Anna Suliga and Marco Drewes for helpful discussions and comments. We thank Hans-Thomas Janka and Daniel Kresse for providing the SN progenitor profiles used in this work. The work of GC is supported by the U.S. Department of Energy under the award number DE-SC0020250 and DE-SC0020262. The works of SH, PH, and IMS are supported by the U.S.~Department of Energy Office of Science under award number DE-SC0020262 and DE-SC0020250. The work of SH is also supported by NSF Grant No.~AST1908960, No.~PHY-1914409 and No.~PHY-2209420; and JSPS KAKENHI Grant Numbers JP22K03630 and JP23H04899. This work was supported by World Premier International Research Center Initiative (WPI Initiative), MEXT, Japan. 

\section*{Note Added}
Two months after our work appeared on the arXiv pre-print server, another work appeared about supernovae constraints on sterile neutrinos~\cite{Carenza:2023old}. In their work, the authors point out that the sterile neutrino production increases drastically by including the scattering of neutrinos off of neutrons and protons, i.e., $\nu+N \rightarrow \nu_s +N$, which had been missed by previous studies. We have verified their result by including this missing production channel and find that the low-energy SN constraint gets stronger by 1-2 orders of magnitude. \GC{In addition, our work delves deeper into the observational aspects for low-energy supernovae constraints and also compares the effect of envelope size, gravitational trapping, visible energy deposition and different progenitor models on the SNIIP constraints.} 

We also find that the bounds in their work are slightly stronger than reported by us in Fig.~\ref{fig:mainplot}. The primary reasons being their assumptions about larger SN core radius\footnote{\GC{Naively this volume enhancement leads to their SNIIP constraints stronger by a factor of 8 compared to a 20 km SN core profile.}} ($40$ km) and larger envelope size ($2.5\times 10^{13}$ cm). However, our assumptions on SN core radius ($20$ km) and envelope size ($5\times 10^{12}$ cm) are more in line with the standard expectation for SN progenitors with lowest explosion energies $\sim 10^{50}$ erg. 

\appendix
\section{Appendix A}
\label{app:A}
\GC{The sterile neutrino production rate through neutrino scattering off nucleons $N$ depends on the following expression $|\mathcal{M}|^2_{VV}+ |\mathcal{M}|^2_{VA}+ |\mathcal{M}|^2_{AA}$~\cite{Bruenn1985,Carenza:2023old}, with
\begin{eqnarray}
   |\mathcal{M}|^2_{VV} &=& 2\, G_V^2\, \left[ (p_1\cdot p_2)(p_3\cdot p_4) + (p_2\cdot p_4)(p_1\cdot p_3) - m_2\,m_3\,(p_1\cdot p_4) \right] \,\ , \label{MVV} \\
  |\mathcal{M}|^2_{VA} &=&  4\,G_V\,G_A\,[(p_1\cdot p_2)(p_3\cdot p_4)-(p_2\cdot p_4)(p_1\cdot p_3)] \,\ , \label{MVA} \\
 |\mathcal{M}|^2_{AA} &=&  2\,G_A^2\,[(p_1\cdot p_2)(p_3\cdot p_4)+(p_2\cdot p_4)(p_1\cdot p_3)+m_2\,m_3\,(p_1\cdot p_4)]]  \,\ , \label{MAA} \\
   G_V&\to& G_V^n=\frac{1}{2} \,\,\,\,\,\, , \,\,\, G_V^p=\frac{1}{2}-2\sin^2\theta_W \,\ , \\
    G_A&\to& G_A^n=\frac{g_A}{2} \,\ , \,\ G_A^p=\frac{g_A}{2} \,\ .
\end{eqnarray}
where the variables $G_{V,A}$ with superscript $(n,p)$ stands for vector and axial charges for neutron and proton respectively, and $g_A=1.27$ the axial vector coupling constant, respectively.}

\bibliography{ref}

\end{document}